\documentclass[conference]{IEEEtran}

\usepackage{graphicx}
\usepackage{amsmath}
\usepackage{caption}
\usepackage{epstopdf}
\usepackage{float}
\usepackage{array}

\begin{document}
%
\title{Reconfigurable Dual Mode IEEE 802.15.4 Digital Baseband Receiver for Diverse IoT Applications }

\author{\IEEEauthorblockN{Mohammed Abdullah Zubair, P. Rajalakshmi}
\IEEEauthorblockN{Department of Electrical Engineering\\
Indian Institute of Technology - Hyderabad, India\\
Email: ee14mtech01003, raji@iith.ac.in\\
}
}
\maketitle

\begin{abstract}
IEEE 802.15.4 takes a center stage in IoT as Low-Rate Wireless Personal Area Networks(LR-WPANs). The standard specifies Offset Quadrature Phase Shift Keying Physical Layer (O-QPSK PHY) with half-sine pulse shaping which can be either analyzed under the class of M-ary PSK signals (QPSK signal with offset) or as Minimum Shift Keying (MSK) signal. M-ary PSK demodulation requires perfect carrier and has minimal error. MSK signals which falls under Continuous Phase Frequency Shift Keying can be demodulated non-coherently but error performance is not as good. In our paper, this dual nature of IEEE 802.15.4 PHY is exploited to propose a dual mode receiver comprising of  QPSK demodulator chain and  MSK demodulator chain as a single system on chip. The mode can be configured manually depending on the type of application or based on the feedback from a Signal to Noise (SNR) indicator employed in the proposed receiver. M-ary PSK chain is selected for lower SNRs and MSK for higher SNRs. Each of these properties are analyzed in detail for both demodulator chains and we go on to prove that MSK detection can be used for low power, low complex and low latency while QPSK detection is employed for minimal error.

\end{abstract}


\IEEEpeerreviewmaketitle

\section{Introduction}

As IoT expands its reach across the various aspects of the physical world, we are slowly witnessing its integration within our daily life. In a short period of time, IoT has found manifold and diverse applications, ranging from health care to entertainment, banking to home automation, indoor and outdoor, mobile and stationery. IoT devices in each of these applications have their own constraints. Healthcare devices, for example, need to have minimum error-tolerance while power consumption may not be a constraint. Environment monitoring devices on other hand can be more error tolerant but should survive on batteries for a long duration. This paper proposes a single system on chip with dual mode for IEEE 802.15.4 receiver. The receiver can be configured manually based on the constraints of the application or from the feedback from the built-in SNR estimator.

IEEE 802.15.4 standard specifies Offset Quadrature Phase Shift Keying Physical Layer (OQPSK PHY) with Direct Sequence Spread Spectrum (DSSS) and half-sine pulse shaping \cite{standard}. OQPSK with half-sine pulse shaping is equivalent to Minimum Shift Keying (MSK) signal \cite{pasupathy}. Thus, IEEE 802.15.4 PHY receiver can also be realized as Continuous Phase Frequency Shift Keying (CPFSK) demodulator. MSK, being a special case of CPFSK, is demodulated non coherently. Thus, this type of demodulator do not need carrier frequency and phase synchronization.

As an alternative, at the receiver, I component of the O-QPSK signal is delayed by half a bit duration to match with Quadrature component and then fed to the M-ary PSK demodulator. The coherent QPSK demodulator, though complex to design has excellent error performance. For an optimum performance, M-ary PSK demodulator should be chosen when channel conditions are not favorable and MSK demodulator for high SNRs. The proposed receiver employs a simple SNR indicator based on the preamble to decide when to switch between the two demodulator chains. The SNR indicator observes the extent to which the received preamble differs from the reference preamble to predict the SNR. 

MSK detectors has been used for IEEE 802.15.4 earlier. \cite{simple_msk} demonstrates a simple low cost receiver using Asynchronous Zero Crossing Detector (AZCD) as a form of MSK detector. A Maximum Likelihood Estimation (MLE) based non-coherent MSK demodulation is studied in \cite{eff_msk}. M-ary PSK demodulators with frequency offset synchronization has also been implemented in \cite{else_qpsk} and \cite{offset_qpsk}. These detectors performs better in terms of error performance than the MSK detectors. A multi mode transceiver \cite{prev_multi} has also been proposed earlier but it is designed for three different modulation schemes for IEEE 802.15.4. In this paper, a dual mode receiver is proposed where in the receiver can switch between demodulator chains and does not need the modulation scheme of the transmitter to change. The only improvisation in the transmitter is the differential encoding of the source bit stream. This is for the reason that differential encoded bits with OQPSK modulation and half sine pulse shaping is exactly equivalent to MSK modudlation. If the QPSK demodulator chain is used at the receiver, a differential decoder is used to retrieve original bit stream. MSK demodulator need not use a differential decoder.

The  paper is organized as follows, the proposed system architecture is discussed in section II. The flow of both the demodulator chain is explained in section III. The simulation results is under section III and Section IV concludes the paper with future scope.

\section{Proposed System Architecture}

\begin{figure}[h]
	\begin{center}
		
		\includegraphics[width = 9cm,height = 7cm]{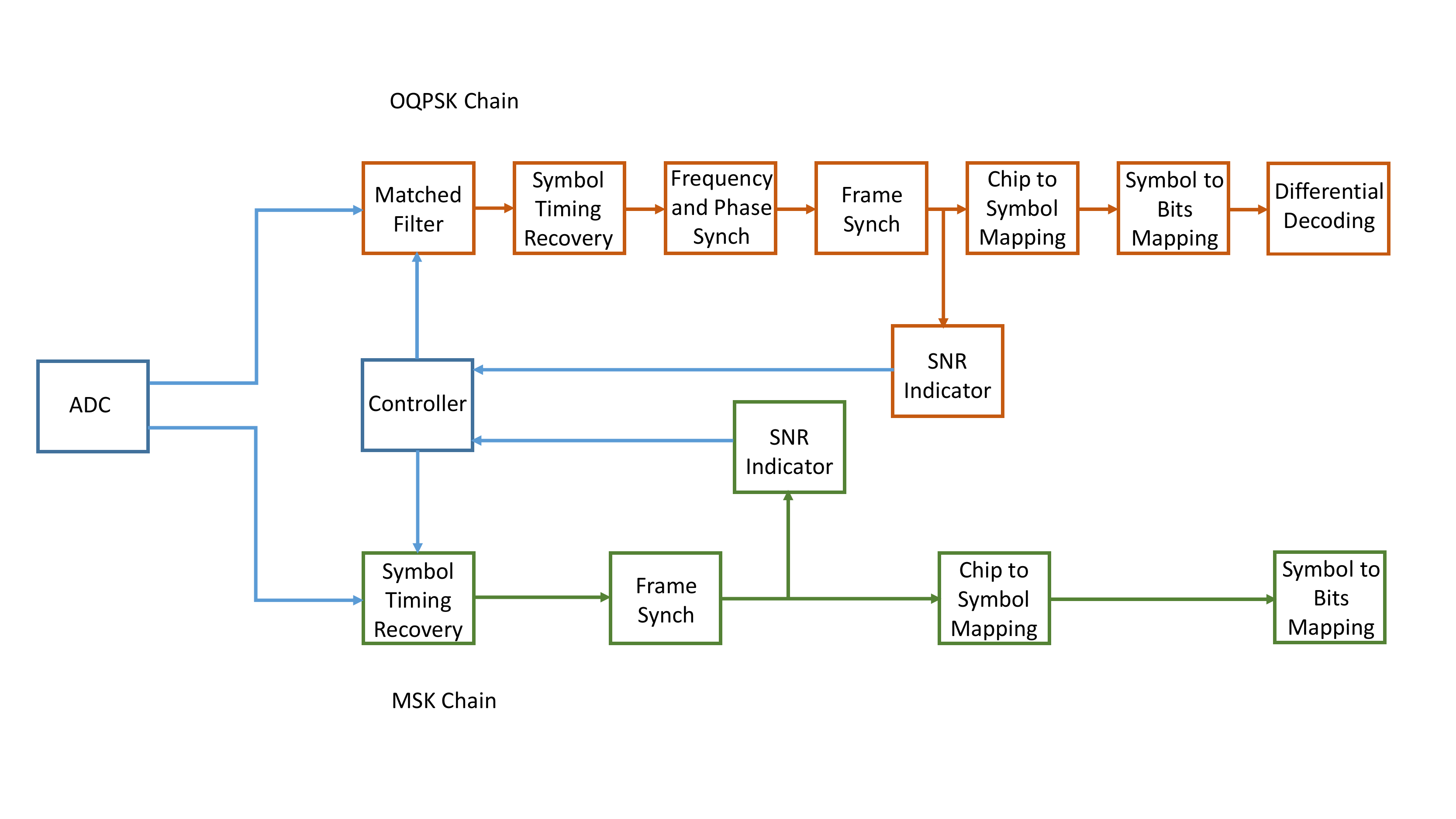}
	\end{center}
	\caption{Proposed IEEE 802.15.4 Dual Mode Receiver}
	\label{fig:sys_arch}
\end{figure}

The complete architecture of the proposed IEEE 802.15.4 receiver is shown in figure \ref{fig:sys_arch}. Each of the chains have their own SNR Estimator that gives feedback to the controller. Based on this feedback, the controller directs the data stream from the ADC to one of the demodulator chains. The digital baseband signal with synchronization errors is given as:
\begin{align}
z(t) = exp(j(2\pi f_dt + \theta))s(t-\tau) + n(t)
\label{main_eq}
\end{align}
where $f_d$ is the frequency offset, $\theta$ is the phase offset, $\tau$ is the timing error, $s(t)$ is the modulated signal and $n(t)$ is Additive White Gaussian Noise (AWGN). As apparent from \eqref{main_eq}, the offsets and timing error should be compensated before decoding and decision.

  The algorithms used in each of the demodulator chains is explained in the following sub sections.

\subsection{IEEE 802.15.4 OQPSK Demodulator}
The offset between In-phase and Quadrature-phase components of the OQPSK signal is removed by delaying the \textit{I} component to align with \textit{Q} component in the Symbol Timing Recovery block. The signal is then be fed to the QPSK demodulator. 
\subsubsection{Matched Filter and Symbol Timing Recovery}
A discrete matched filter, $h[n]$ with coefficients equivalent to that of half-sine pulse shape is used to smoothen the pulse of both I and Q signals.
\begin{align}
h[n] = sin(\frac{\pi n}{NT_s})
\label{half_sine}
\end{align}
Since, the half-sine pulse is symmetric, $h^\ast[-n] = h[n]$.

\begin{figure}[h]
	\begin{center}
		
		\includegraphics[scale=0.27]{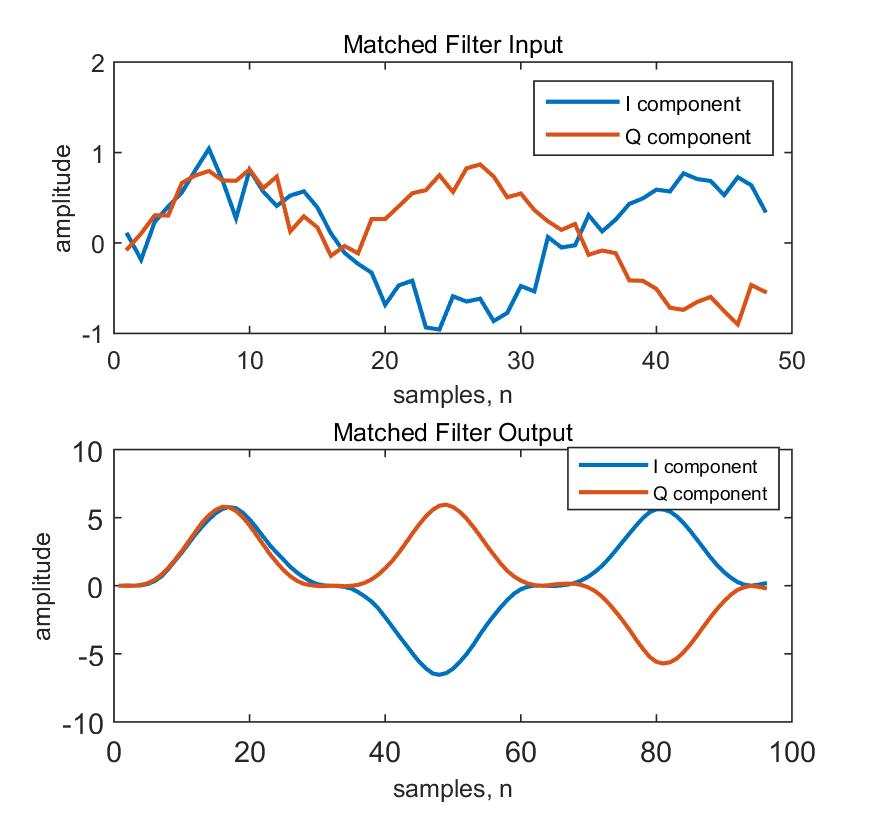}
	\end{center}
	\caption{Matched Filter Operation}
	\label{fig:mf}
\end{figure}

\begin{figure}[h]
 	\begin{center}
 		
 		\includegraphics[scale=0.33]{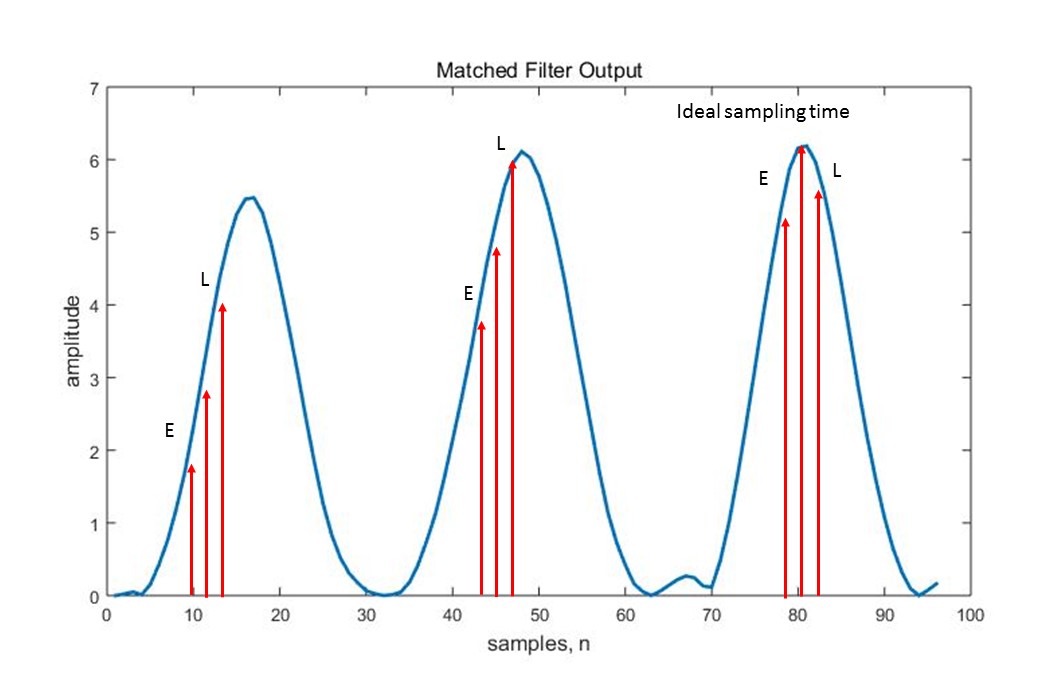}
 	\end{center}
 	\caption{Illustration of Early Late Gate Algorithm}
 	\label{fig:elg}
 \end{figure}

 Early Late Gate (ELG) Algorithm \cite{proakis} is then applied for symbol timing recovery. ELG is a feedback synchronizer and is usually carried on for a fixed number of pulses till an approximate timing interval is achieved. Two additional samples with the main sample is taken, namely early sample and late sample, E and L as shown in figure \ref{fig:elg}. The samples are shifted towards left or right in the next pulse till the main sample is found to be of greater amplitude than both E and L. If L is greater, samples are shifted right, else towards left. 
 
\subsubsection{Carrier Frequency and Phase Synchronizer} 
An excellent tutorial of Digital frequency offset estimators for M-ary PSK modulation is given in \cite{freq_est_tutorial}. Both the Data Aided (DA) and Non Data Aided (NDA) estimators are explained. We have used NDA estimator because the frequency offset estimation is performed before frame synchronization in our receiver and hence preamble cannot be used before that. As discussed in the tutorial, R and B algorithm \cite{RB_freq} is the most powerful estimator even at low SNRs. Log likelihood function of NDA model of R and B algorithm is given by :

\begin{align}
\Lambda_l = e^{-j\theta}\ \sum_{n=0}^{N_{fft}-1}z[n]e^{-j2\pi nf_dT}\
\label{llf}
\end{align} 
where, $z[n] = e^{jM(2\pi nf_dT+\theta+\eta_n)}$\\
$M = 4$ in case of OQPSK and $\eta_n$ is the phase of Gaussian noise, as stated in \cite{freq_est_tutorial}\\
The estimates of frequency offset and phase offset is then evaluated as:

\begin{align}
\hat{f_d}= max\Big\{\Big\lvert \sum_{n=0}^{N_{fft}-1}z[n]e^{-j2\pi nf_dT}  \Big\rvert \Big\} \\
\hat{\theta} = max\Big\{e^{-j\theta} \Delta(\hat{f_d})\}
\label{estimators}
\end{align} 

where, $\Delta(\hat{f_d}) =  \sum_{n=0}^{N_{fft}-1}z[n]e^{-j2\pi n\hat{f_d}T}  $

\begin{figure}[h]
	\begin{center}
		
		\includegraphics[scale=0.38]{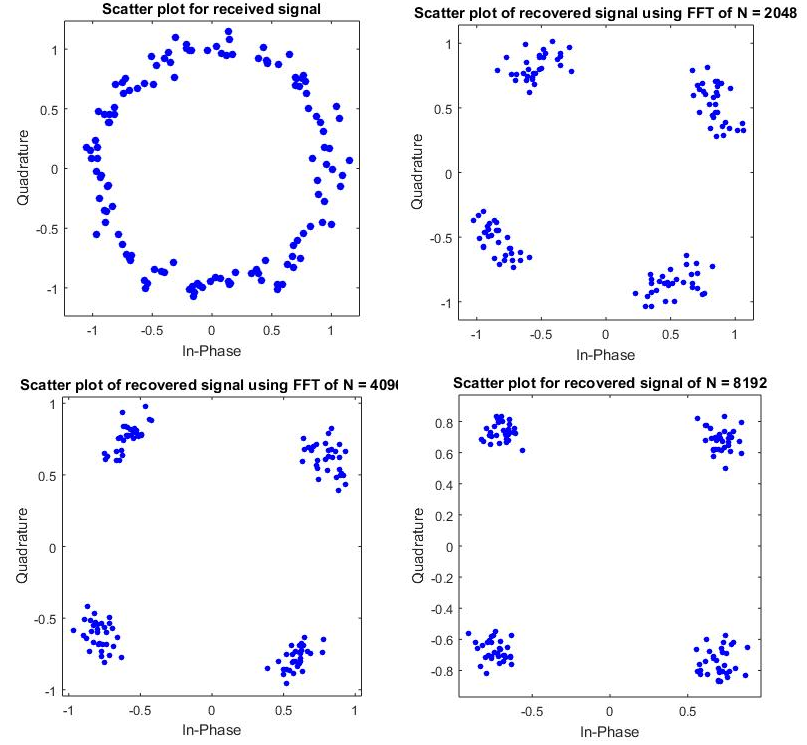}
	\end{center}
	\caption{Frequency and phase offset compensation for increasing values of $N_{fft}$}
	\label{fig:fdtheta}
\end{figure}

The performance of the estimator increases with high values of $N_{fft}$ as shown in figure \ref{fig:fdtheta}.

\subsubsection{Frame Synchronization}
As per the standard, IEEE 802.15.4 PHY layer packet has 32 zeros as a preamble. This 32 length preamble is converted to 8 symbols from bit to symbol mapping. Each of these symbols is then mapped to 32 bit chip sequence by DSSS, thus giving us a known preamble sequence of 256 bits or 128 QPSK symbols. A 128 symbol correlator is used to detect the preamble and synchronize the frame. The payload is then extracted and sent to \textit{chip to symbol mapping} and then to \textit{symbol to bit mapping}.

\subsection{IEEE 802.15.4 MSK Demodulator}
\begin{figure}[h]
	\begin{center}
		
		\includegraphics[scale=0.27]{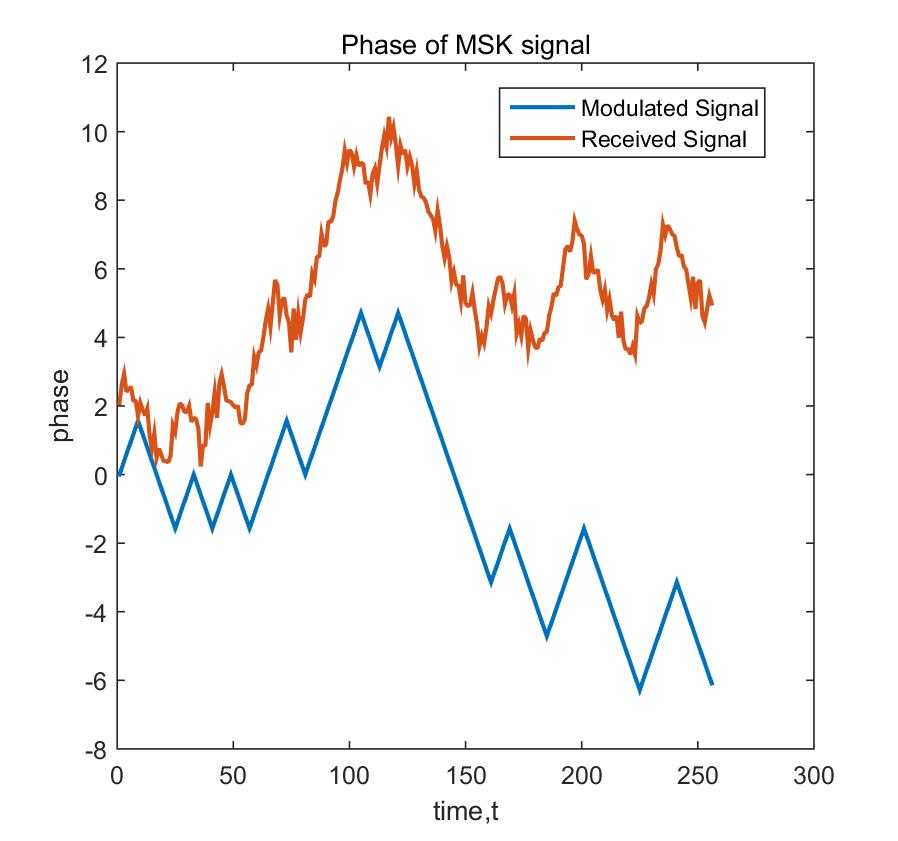}
	\end{center}
	\caption{Continuous phase of MSK}
	\label{fig:msk_phase}
\end{figure}

 The MSK form of IEEE 802.15.4 baseband signal can be considered as : 
 
\begin{align}
z(t) = e^{j(\phi(t-\tau) + 2\pi f_dt + \theta)}\\
\phi(t) = 2\pi h\sum_{k}b_kq(t-kT)
\label{msk_eq}
\end{align}
$h = 0.5$ for MSK, $b_k$ are symbols {-1,1} derived from bits from I and Q components of OQPSK signal and $q(t)$ is  given by:  
\begin{align}
q(t)=
\begin{cases}
0,\ \ \ \ \ t<0 \\
t/2T,\ \ 0<t<T \\
1/2,\ \ \ t>T
\end{cases}
\end{align}
 The continuous phase of MSK signal, $\phi(t)$ can be seen in \ref{fig:msk_phase}. The noisy received version is also shown. As discussed extensively in \cite{pasupathy}, the phase changes by $\pi/2$ or $-\pi/2$ for every transition of $b_k$.

\subsubsection{Symbol Timing Recovery}
The algorithm proposed in \cite{str_msk} is used for timing correction. The non-linear transformation of received signal is taken and then averaged over large number of samples to evaluate the timing error, $\tau$.
\begin{align}
c_{n,i} = [z_{n,i}z^*_{n-1,i}]^2 \\
v_{n,i} = E\{c_{n,i}\}
\end{align}

The minimum timing error will occur at that value of $i$ which gives maximum $|v_{n,i}|$
\begin{align}
\hat{\tau}=\underset{i}{\max}\{|v_{n,i}|\}
\end{align}
where $z_n$ is the discrete version of \eqref{msk_eq} and $n$ is $n^{th}$ symbol and $i$ is the number of samples in one symbol.

\subsubsection{MSK Detection}
\begin{figure}[h]
	\begin{center}
		
		\includegraphics[scale=0.29]{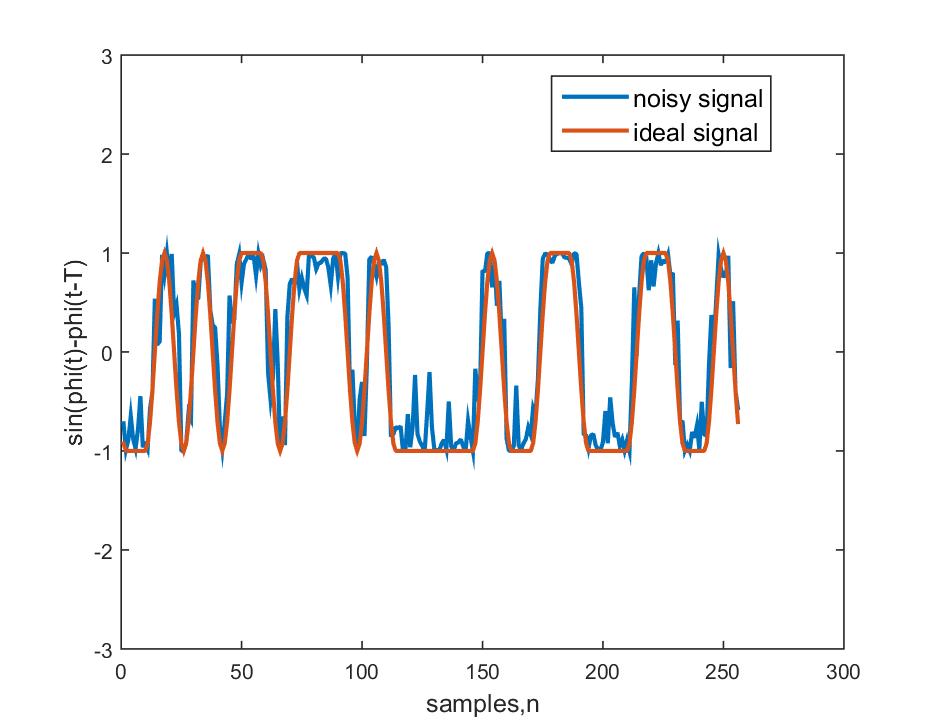}
	\end{center}
	\caption{Differential Detection of MSK}
	\label{fig:sine_phase}
\end{figure}

Differential Detection of MSK is performed to eliminate the effects of frequency and phase offsets. This approach was given by Tatsuro et. $al$ in \cite{tatsuro}. The sine of difference of $\phi(t)$ and $\phi(t-T)$ is evaluated and decision is made by checking the sample just before each transition. If the sample is above 0, the bit is taken to be 1, else it is taken to be 0. This is shown in figure \ref{fig:sine_phase}. The bit stream is then sent to \textit{chip to symbol} and \textit{symbol to bit} mapping.

\subsection{SNR Indicator}
As a novel approach, Frame Synchronizer can be used as a crude indicator of SNR. Both the receiver chains need to match the chip sequence corresponding to 32 zeros for preamble detection. After the preamble is detected, a 256-bit comparator can be employed to match the received preamble sequence with original preamble sequence. A higher SNR will yield in more number of matching bits while lower SNRs degrade the comparator output. A threshold is set, output above the threshold indicates a good SNR and output below the threshold indicates a lower SNRs. QPSK demodulator is selected for lower SNRs and MSK demodulator for good SNRs. While one chain is processing, the other will sleep. This adaptive switching between the demodulator chains optimizes the performance of the receiver.

\begin{table*}[!htp]
	\begin{center}
		\begin{tabular}{ | m{5em} | m{10em} | m{3cm}| m{3cm} | m{3cm} | } 
			\hline
			\textbf{Algorithm} & \textbf{Mathematical Functions} & \textbf{Additions} & \textbf{Multiplications} & \textbf{Other Operations} \\  
			\hline
			Symbol Timing Recovery & Complex Multiplications \& Complex number squaring & $L*N_{sample}/2*4$ & $L*N_{sample}/2*8$  & $N_{sample}/2 - 1$ \\ 
			\hline
			Frame Synchronization & Real Cross Correlation & $(N_{preamble}-1)^2$ & $(N_{preamble})^2$ & Nil \\ 
			\hline
			Chip to Symbol Mapping & XOR Operation & $(N_{bits}/4)*31*16$ & Nil & $N_{bits} * 128$ XOR \& $(N_{bits}/4)*15$ comparisons  \\
			\hline
			Bit Detection & Nil & Nil & Nil & $N_{bits}$ comparisons  \\
			\hline
		\end{tabular}
		\caption{Complexity Analysis of MSK Demodulator}
		\label{msk_table}
	\end{center}
\end{table*}

\begin{table*}[!htp]
	\begin{center}
		\begin{tabular}{ | m{5em} | m{10em} | m{4cm}| m{4cm} | m{3cm} | } 
			\hline
			\textbf{Algorithm} & \textbf{Mathematical Functions} & \textbf{Additions} & \textbf{Multiplications} & \textbf{Other Operations} \\ 
			\hline
			Matched Filter and Symbol Timing Recovery & Convolution  & $L*2*(N_{sample}-1)^2$  & $L*2*(N_{sample})^2$ & $L*3$ comparisons \\ 
			\hline
			Frequency and Phase Synchronization & Fast Fourier Transform \footnotemark & $7\frac{N_{fft}}{2}\log(\frac{N_{fft}}{2})-5N_{fft}+8$ \cite{fft_radix}& $3\frac{N_{fft}}{2}\log(\frac{N_{fft}}{2})-5N_{fft}+8$ \cite{fft_radix} & $1*arc tan$ \\
			\hline
			Frame Synchronization & Real Cross Correlation & $(N_{preamble}-1)^2$ & $(N_{preamble})^2$ & Nil \\ 
			\hline
			Chip to Symbol Mapping & XOR Operation & $(N_{bits}/4)*31*16$ & Nil & $N_{bits} * 128$ XOR \& $(N_{bits}/4)*15$ comparisons  \\
			\hline
			Bit Detection & Nil & Nil & Nil & $N_{bits}$ comparisons  \\
			\hline
		\end{tabular}
		\caption{Complexity Analysis of QPSK Demodulator}
		\label{qpsk_table}
	\end{center}
\end{table*}

\section{Performance Analysis and Simulation Results}

\subsection{Bit Error Rate Performance}

\begin{figure}[h]
	\begin{center}
		
		\includegraphics[scale=0.45]{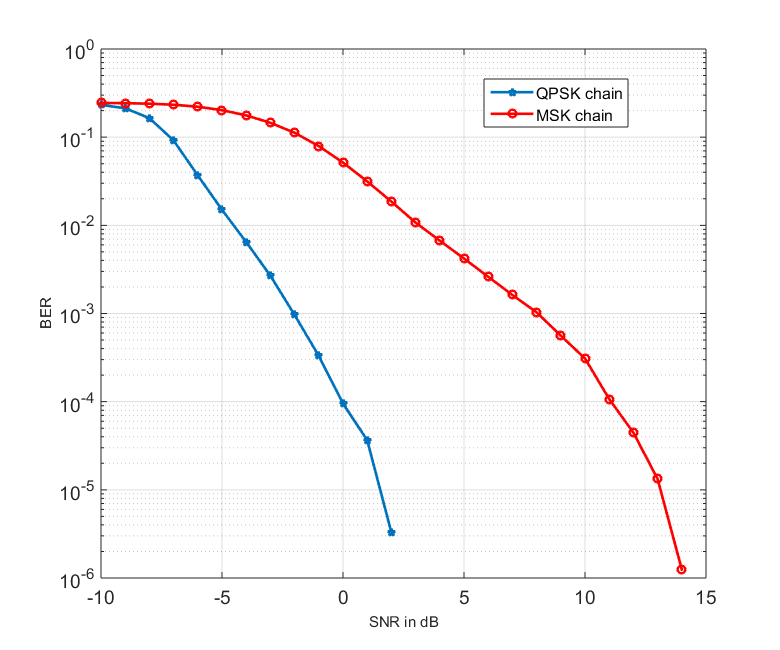}
	\end{center}
	\caption{BER performance of both the demodulators}
	\label{fig:ber_both}
\end{figure}

As shown in figure \ref{fig:ber_both}, the QPSK demodulator clearly outperforms MSK demodulator. At bit error rate of $10^{-3}$, there is a huge difference of 16 dB between the two. High BER till $-10 dB$ is observed due to the inability of demodulators to detect the preamble and synchronize the frame. Hence no further processing takes place and the packet is missed.

\subsection{Complexity Analysis}
The computational complexity of algorithms used in both the demodulator chains is given in the tables \ref{msk_table} and \ref{qpsk_table}.\\
 $L$ is the number of symbol pulses used for symbol timing recovery, $N_{sample}$ is the number of samples per symbol pulse, $N_{preamble}$ is the number of preamble symbols considered, maximum of $N_{preamble}$ is 128 for QPSK and 256 for MSK, $N_{bits}$ is the length of the transmitted bit stream and $N_{fft}$ is the length of FFT performed in frequency offset estimation of QPSK signal.\\
 As it is apparent from the tables that QPSK chain has higher complexity. For the values, $L=32$,  $N_{sample}=16$, $N_{preamble}=64$, $N_{bits}=200$ and $N_{fft}=2048$, MSK demodulator needs 29,793 additions and 6,144 multiplications while QPSK demodulator requires 104,617 additions and 40,695 multiplications.
 
 Also, the operations like FFT and Matched Filter can't be implemented on-the-fly and needs sufficient memory for storage and processing. The read and write operations of the memory introduces large latency in the system. Power consumption also rises with usage of memory and more number of adders and multipliers.
 
 Thus, MSK chain can be termed as low complex, low power and low latency demodulator while the QPSK chain can be seen as highly efficient and minimal error demodulator.

\footnotetext{Cooley and Tukey Algorithm with radix-2}

\section{Conclusion and Future Scope}
We have proposed a dual mode receiver for IEEE 802.15.4 and have shown that it can adapt to any application. The MSK detector can be selected for applications that needs to be in the network for a long duration and can compensate for few errors. Applications where latency is also a constraint can use this chain. On the other hand, QPSK detector gives efficient performance in terms of bit error rate. IoT applications where errors cannot be tolerated can be operated in this mode. Also, the receiver itself can switch between the modes using the SNR Indicator to optimize the performance and balance between power consumption, latency and error performance. The proposed receiver design can be implemented as a single system on chip and can deliver optimum performance for universal IoT applications.


\begin{thebibliography}{1}
\bibitem{standard}
IEEE Standard for Local and metropolitan area networks--Part 15.4: Low-Rate Wireless Personal Area Networks (LR-WPANs), IEEE 802.15.4™-2011.
\bibitem{pasupathy}
S. Pasupathy, "Minimum shift keying: A spectrally efficient modulation," in IEEE Communications Magazine, vol. 17, no. 4, pp. 14-22, July 1979.
\bibitem{simple_msk}
N. Dehaese, S. Bourdel, H. Barthelemy and G. Bas, "Simple demodulator for 802.15.4 low-cost receivers," 2006 IEEE Radio and Wireless Symposium, 2006, pp. 315-318.
doi: 10.1109/RWS.2006.1615158
\bibitem{eff_msk}
Shouyi Yin, Jianwei Cui, Ao Luo, Leibo Liu and Shaojun Wei, "A high efficient baseband transceiver for IEEE 802.15.4 LR-WPAN systems," ASIC (ASICON), 2011 IEEE 9th International Conference on, Xiamen, 2011, pp. 224-227.
doi: 10.1109/ASICON.2011.6157162
\bibitem{else_qpsk}
Shengchen Dai.; Hua Qian .;·Kai Kang ·.;Weidong Xiang.
"A Robust Demodulator for OQPSK–DSSS System",
Circuits Syst Signal Process (2015) 34:231–247
DOI 10.1007/s00034-014-9844-z
\bibitem{offset_qpsk}
D. Park, C. S. Park and K. Lee, "Simple Design of Detector in the Presence of Frequency Offset for IEEE 802.15.4 LR-WPANs," in IEEE Transactions on Circuits and Systems II: Express Briefs, vol. 56, no. 4, pp. 330-334, April 2009.
doi: 10.1109/TCSII.2009.2015384
\bibitem{prev_multi}
P. Wu, C. Zhang, C. Wei, H. Jiang and Z. Wang, "A baseband transceiver for multi-mode and multi-band SoC," 2012 IEEE 55th International Midwest Symposium on Circuits and Systems (MWSCAS), Boise, ID, 2012, pp. 770-773.
doi: 10.1109/MWSCAS.2012.6292134
\bibitem{proakis}
J. G. Proakis., Masoud Salehi., Digital Communications. McGraw Hill.Fifth Edition.
\bibitem{freq_est_tutorial}
Morelli, M. and Mengali, U. (1998), Feedforward frequency estimation for PSK: A tutorial review. Eur. Trans. Telecomm., 9: 103–116. doi: 10.1002/ett.4460090203
\bibitem{RB_freq}
D. Rife and R. Boorstyn, "Single tone parameter estimation from discrete-time observations," in IEEE Transactions on Information Theory, vol. 20, no. 5, pp. 591-598, Sep 1974.
doi: 10.1109/TIT.1974.1055282
\bibitem{str_msk}
T. Masamura, S. Samejima, Y. Morihiro and H. Fuketa, "Differential Detection of MSK with Nonredundant Error Correction," in IEEE Transactions on Communications, vol. 27, no. 6, pp. 912-918, Jun 1979.
doi: 10.1109/TCOM.1979.1094478
\bibitem{tatsuro}
T. Masamura, S. Samejima, Y. Morihiro and H. Fuketa, "Differential Detection of MSK with Nonredundant Error Correction," in IEEE Transactions on Communications, vol. 27, no. 6, pp. 912-918, Jun 1979.
doi: 10.1109/TCOM.1979.1094478
\bibitem{fft_radix}
	P. Duhamel,	M. Vetterli, "Fast fourier transforms: a tutorial review and a state of the art," in Journal,Signal Processing archive,Volume 19 Issue 4, April 1990 Pages 259 - 299 


\end{thebibliography}
\end{document}